\documentclass[a4paper,11pt]{article}
\usepackage{pos}
\usepackage{color}
\usepackage{subfigure}
\graphicspath{ {image/} }
\usepackage{multirow}
\newcommand*\diff{\mathop{}\!\mathrm{d}}
\usepackage{braket}
\usepackage{bm}
\newcommand*\GeV{\mathop{}\!\mathrm{GeV}}

\title{Forward quark jet-nucleus scattering in a light-front Hamiltonian approach}

\author*[a,b,c]{Meijian Li}

\affiliation[a]{Department of Physics, University of Jyv\"{a}skyl\"{a}, \\
P.O. Box 35, FI-40014 
Finland}
\affiliation[b]{
Helsinki Institute of Physics, University of Helsinki,\\ P.O. Box 64, FI-00014,
Finland
}
\affiliation[c]{Department of Physics and Astronomy, Iowa State University,\\ Ames, IA, 50011, USA}

\emailAdd{mliy@jyu.fi}

\abstract{
  We investigate the scattering of a quark jet on a high-energy heavy nucleus using the time-dependent light-front Hamiltonian approach. 
  We simulate a real-time evolution of the quark in a strong classical color field of the relativistic nucleus, described as the Color Glass Condensate. 
  We study the sub-eikonal effect by letting the quark jet carry realistic finite longitudinal momenta, and we find sizeable changes on the transverse coordinate distribution of the quark. We also observe the energy loss of the quark through gluon emissions in the $\ket{q}+\ket{qg}$ Fock space.
  This approach provides us with an opportunity to study scattering processes from non-perturbative aspects.
}

\FullConference{%
  HardProbes2020\\
  1-6 June 2020\\
  Austin, Texas}

\begin{document}
\maketitle

\section{Introduction}
Scattering of an ultrarelativistic quark off a heavy nucleus is one of the most direct ways to study the structure of the cold nuclear matter at low values of Bjorken’s $x$. 
In a recent work~\cite{Li:2020uhl}, we investigated the sub-eikonal non-perturbative corrections to the quark-nucleus scattering using a light-front Hamiltonian formalism, the time-dependent basis light-front quantization (tBLFQ)~\cite{Zhao:2013cma}. 
In this proceeding, we extend the investigation by including one dynamical gluon.

\section{Time-dependent basis light-front quantization (tBLFQ)}\label{sec:method}
We consider scattering of a high-energy quark moving in the positive $z$ direction, on a high-energy nucleus moving in the negative $z$ direction. The quark has momentum $p^\mu$ and $p^+>>p^-, p_\perp$ whereas the nucleus has momentum $P^\mu$ and $P^->>P^+, P_\perp$. We treat the quark state at the amplitude level and the nucleus as an external background field. 
The quark interacts with the nuclear field for a finite width along $x^+$. 

The QCD Lagrangian with a background gluon field reads, 
\begin{align}\label{eq:Lag}
  \mathcal{L}=-\frac{1}{4}{F^{\mu\nu}}_a F^a_{\mu\nu}+\overline{\Psi}(i\gamma^\mu \bm D_\mu - \bm m)\Psi\;,
\end{align}
where $D^\mu\equiv \partial_\mu \bm I+ig \bm{\mathcal{A}}^\mu$, $\bm m = m_q \bm I$ ($m_q$ is the quark mass) and $\bm I$ is the 3 by 3 unit matrix in color space. 
$F^{\mu\nu}_a\equiv\partial^\mu C^\nu_a-\partial^\nu C^\mu_a-gf^{abc}C^\mu_b C^\nu_c$ is the field tensor, and $D^\mu\equiv \partial_\mu \bm I+ig\bm C^\mu$.
$\bm C^\mu=\bm A^\mu + \bm{\mathcal{A}}^\mu$ is the summation of the quantum gauge field $\bm A^\mu= A^{a \mu} T^a$ and the background gluon field $\bm{\mathcal{A}}^\mu = \mathcal{A}^{a \mu} T^a$.
The background field is described by the CGC formalism~\cite{Jalilian-Marian:2017ttv}.  
The local density of charge charge in the nucleus is treated as a stochastic variable satisfying the correlation relation,
\begin{equation}\label{eq:chgcor}
  \Braket{\rho_a(\vec{x}_\perp,x^+)\rho_b(\vec{y}_\perp,y^+)}=g^2\tilde{\mu}^2\delta_{ab}\delta^2(\vec{x}_\perp-\vec{y}_\perp)\delta(x^+-y^+)\;.
\end{equation}
The field is solved from \(
  (m_g^2-\nabla^2_\perp )  \mathcal{A}^-_a(\vec{x}_\perp,x^+)=\rho_a(\vec{x}_\perp,x^+)
  \) in the covariant gauge of $\partial^\mu \mathcal{A}_\mu = 0$ and it has only one nonzero component $\mathcal{A}^-$.

The light-front Hamiltonian can be derived from the Lagrangian through the standard Legendre transformation~\cite{Brodsky:1997de}. We write it into two parts as $P^-(x^+) = P_{KE}^- + V(x^+)$. $P_{KE}^-$ is the kinetic energy of the quark and the dynamical gluon. $V(x^+)$ are the remaining interaction terms, which in general, could have a time dependence arising from the external field.
The quark state obeys the evolution equation on the light front, which in the interaction picture reads,
\begin{align}
  \label{eq:ShrodingerEq}
  i\frac{\partial}{\partial x^+}\ket{\psi;x^+}_I=\frac{1}{2}V_I(x^+)\ket{\psi;x^+}_I\;,
\end{align}
where $V_I(x^+)=e^{i\frac{1}{2}P^-_{KE}x^+}V(x^+)e^{-i\frac{1}{2}P^-_{KE}x^+}$ is the interaction Hamiltonian in the interaction picture. 
We solve Eq.~\eqref{eq:ShrodingerEq} by decomposing the evolution time $x^+$ into $n$ steps of size $\delta x^+ \equiv x^+/n$,
\begin{align}\label{eq:ShrodingerEqSol}
  \begin{split}
    \ket{\psi;x^+}_I= & \mathcal{T}_+\exp\left[-\frac{i}{2}\int_0^{x^+}\diff z^+V_I(z^+)\right]\ket{\psi;0}_I =\mathcal{T}_+ \lim_{n\to\infty}\prod^n_{k=1}
    \left[1-\frac{i}{2}V_I(x_k^+) \frac{x^+}{n}\right]
    \ket{\psi;0}_I
    \\
  &=\lim_{n\to\infty}\left[1-\frac{i}{2}V_I(x_n^+)\delta x^+\right]\ldots \left[1-\frac{i}{2}V_I(x_1^+)\delta x^+\right]
  \ket{\psi;0}_I
  \;,
  \end{split}
\end{align}
where $\mathcal{T}_+$ is the light-front time ordering. 
This product expansion is exact in the limit of $n\to\infty$. 

The numerical calculation is carried out on the sites of a 3-dimensional discrete space.
The 2-dimensional transverse space is a lattice extending from $-L_\perp$ to $L_\perp$ for each side, with periodic boundary conditions. 
There are a number of $2N_\perp$ lattice sites on each side, with indices $n_x, n_y=-N_\perp,-N_\perp+1,\ldots,N_\perp-1$. The lattice spacing is $a=L_\perp/N_\perp$. 
The corresponding momentum space is also in a periodic lattice extending from $-\pi/a$ to $\pi/a$ with spacing $d_p\equiv \pi/L_\perp$. 
The background field extends from $0$ to $L_\eta$ along $x^+$, and it is discretized into a number of $N_\eta$ layers. The layer index is $k = 1, 2, \ldots, N_\eta$, and each layer has a length of $\tau=L_\eta/N_\eta$. 
In this discretized space, the correlation relation of the color charge as defined in Eq.~\eqref{eq:chgcor} also takes a discrete form as,
\begin{equation}\label{eq:chgcor_dis}
  \begin{split}
  \Braket{\rho_a(n_x,n_y,k)\rho_b(n'_x,n'_y,k')} 
  =g^2\tilde{\mu}^2\delta_{ab}
  \frac{\delta_{n_x,n'_x}\delta_{n_y,n'_y}}{a^2}
  \frac{\delta_{k,k'}}{\tau}\;.
\end{split}
\end{equation}

\section{Calculation in the Fock space of $\ket{q}$ }
To start with, we truncate the Fock space of the quark to the leading sector as $\ket{q}$. In this case, the QCD Lagrangian in Eq.~\eqref{eq:Lag} reduces to \( \mathcal{L}_{q} = \overline{\Psi}(i\gamma^\mu \bm D_\mu - \bm m)\Psi \). The interaction term in the Hamiltonian is between the background field and the quark, 
\begin{align}
P_{KE}^-=\int\diff x^-\diff^2 x_\perp
\frac{1}{2}\overline{\Psi}\gamma^+\frac{m^2-\nabla_\perp^2}{2i\partial_-}\Psi\;,
\qquad
V(x^+)=\int\diff x^-\diff^2 x_\perp
g\bar{\Psi}\gamma^+ T^a\Psi\mathcal{A}^a_+
\;.
\end{align}
Considering that the background field interacts with the quark in the transverse space and the color space, we construct the basis state as $\ket{\beta} = \ket{k_x, k_y, c}$, where $k_x$ and $k_y$ are the transverse momentum of the quark and $c$ is the color of the quark. The quark state is then expanded as
\[ \ket{\psi;x^+}_I=\sum\nolimits_\beta c_\beta (x^+)\ket{\beta}\;,\]
where $c_\beta(x^+)\equiv\braket{\beta|\psi;x^+}_I$ are the basis coefficients. 
The scattering of the quark on a CGC field is simulated according to the non-perturbative method explained in Section~\ref{sec:method}.

\begin{figure*}[htp!]
  \centering
  \subfigure[\ Evolution of the quark's transverse coordinate distribution at $p^+=\infty$  \label{fig:rt2D_evl_pinf}]
  {\includegraphics[width=\textwidth]
  {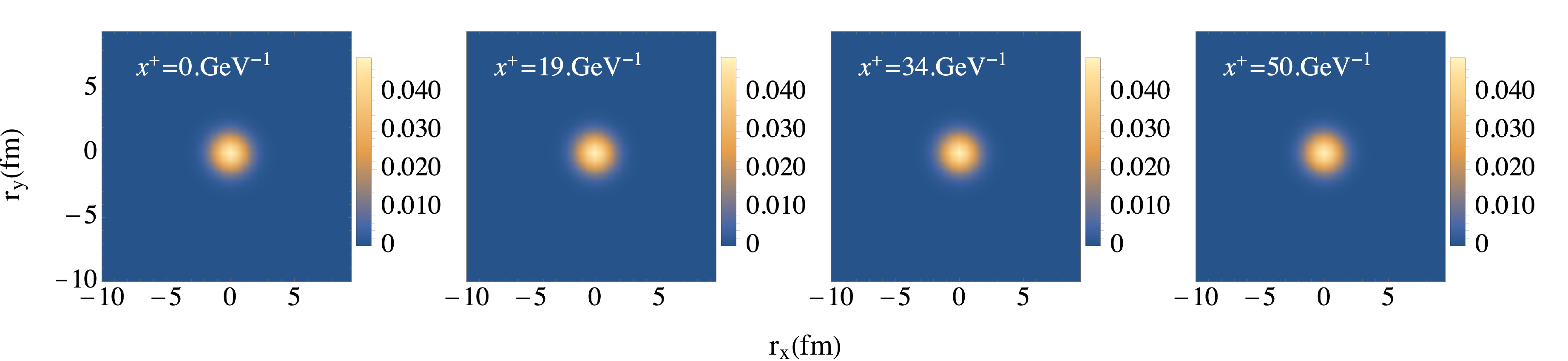}
  }
  \subfigure[\ Evolution of the quark's transverse coordinate distribution at $p^+=10~\GeV$ \label{fig:rt2D_evl_p10_50event}]
  {\includegraphics[width=\textwidth]{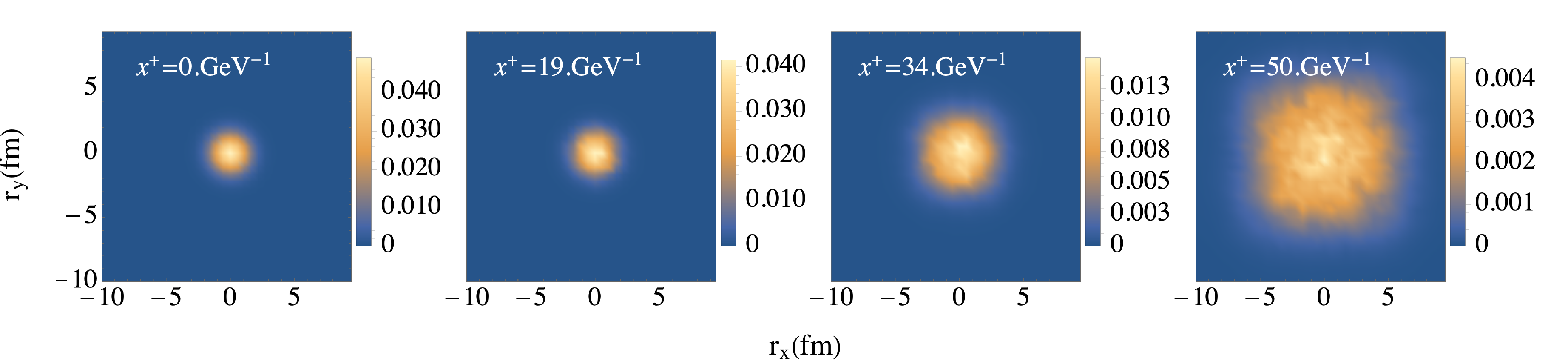}
  }
  \caption{The evolution of the quark's transverse coordinate distribution at different $p^+$, (a) $p^+=\infty$, (b) $p^+=10~\GeV$. 
  The initial state of the quark is distributed as $C e^{-|\vec r_\perp|^2/(0.2*50~\GeV^{-1})^2}$, where $C$ is the normalization coefficient. From left to right, the transverse coordinate distributions of the quark are shown at a sequential interaction time. 
  Parameters in those panels: $L_\eta=50~\GeV^{-1}$, $N_\eta=4$, $m_g=0.1~\GeV$, $N_\perp=18$, $L_\perp=50~\GeV^{-1}$, $g^2\tilde{\mu}=0.486~\GeV^{-3/2}$. 
  (Figure adapted from Ref.~\cite{Li:2020uhl})}
  \label{fig:rt2D_evl}
\end{figure*}

We find that the resulting quark's cross sections agree with the analytical eikonal expectations not only in the eikonal limit of $p^+=\infty$, but also in cases with very small $p^+$~\cite{Li:2020uhl}.
The non-eikonal effect is observed in the transverse coordinate space. As shown in Fig.~\ref{fig:rt2D_evl}, in the eikonal limit of $p^+ =\infty$,  the quark does not change its transverse location. 
In the case with a finite value of $p^+$, the quark spreads out in its transverse coordinate distribution. 

\section{Calculation in the Fock space of $\ket{q} + \ket{qg}$}
In the $\ket{q} + \ket{qg}$ Fock space, the QCD Lagrangian in Eq.~\eqref{eq:Lag} is restored. 
The interaction part contains, the interaction of the external field with the quark, that between the dynamical gluon and the quark, and that of the external field with the dynamical gluon. The instantaneous terms in the $\ket{qg}$ sector are excluded by the ``gauge cutoff'' formulation~\cite{Tang:1991rc}.
\begin{align}
  \begin{split}
    P_{KE}^-=&\int\diff x^-\diff^2 x_\perp
    \bigg\{
    -\frac{1}{2}A^j_a{(i\nabla)}^2_\perp A_j^a
    +\frac{1}{2}\overline{\Psi}\gamma^+\frac{m^2-\nabla_\perp^2}{2i\partial_-}\Psi   \bigg\}\;,\\
    V(x^+)=&\int\diff x^-\diff^2 x_\perp
    \bigg\{
+ g\bar{\Psi}\gamma^+ T^a\Psi\mathcal{A}^a_+
+ g\bar{\Psi}\gamma^\mu T^a\Psi A^a_\mu
+g f^{abc} \partial^+ A^i_b A^c_i \mathcal{A}^a_+
   \bigg\}
   \;.
  \end{split}
\end{align}

The quark state is expanded on the discretized momentum basis as 
\[  \ket{\psi;x^+}_I = \sum\nolimits_{\beta_q} c_{\beta_q} (x^+)\ket{\beta_q} + \sum\nolimits_{\beta_{qg}} c_{\beta_{qg}} (x^+)\ket{\beta_{qg}}\;.\]
$c_{\beta_q}(x^+)\equiv\braket{\beta_q|\psi;x^+}_I$ and $c_{\beta_{qg}}(x^+)\equiv\braket{\beta_{qg}|\psi;x^+}_I$ are the basis coefficients. 
The basis states are 
\(\ket{\beta_q} =\ket{k^x, k^y, k^+, s, c} \) and 
\(\ket{\beta_{qg}} =\ket{k_q^x, k_q^y, k_q^+, s_q, c_q, k_g^x, k_g^y, k_g^+, s_g, c_g} \), for the $\ket{q}$ and the $\ket{qg}$ sector respectively. Compared with bare quark basis $\ket{\beta}$, these basis states also specify the spin projection and the longitudinal momentum of the partons, since these quantum numbers could change via the gluon emission/absorption. The $x^-$ direction is treated as a circle of length $2L$, 
and the longitudinal momentum $p^+$ in the basis states takes the discrete values \(p^+ = (2\pi/L) k^+ \),
where $k^+ = 1, 2, 3,\ldots$ for bosons  (neglecting the zero mode) and $k^+ = 1/2, 3/2,5/2,\ldots$ for fermions. 
We define $K_{\max}$ as the total $k^+$ of the system, and it remains the same during the evolution.

\begin{figure}[htp!]
  \centering
  \subfigure[\ Evolution of $p^+$ in the $\ket{q}$ sector\label{fig:ppl_q}]
  {\includegraphics[width=0.45\textwidth]{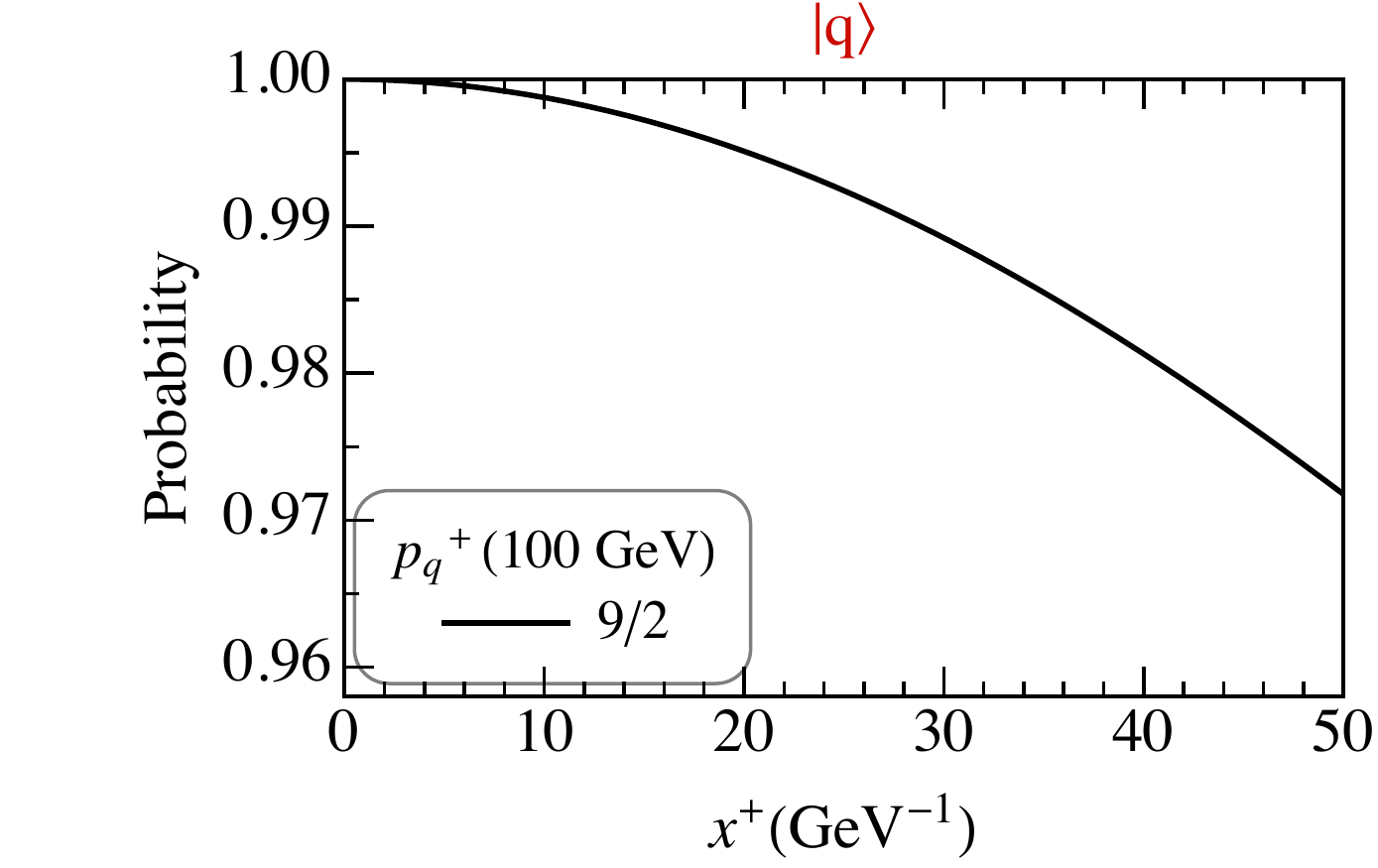}
  } 
  \subfigure[\ Evolution of $p^+$ in the $\ket{qg}$ sector \label{fig:ppl_qg}]
  {\includegraphics[width=0.45\textwidth]{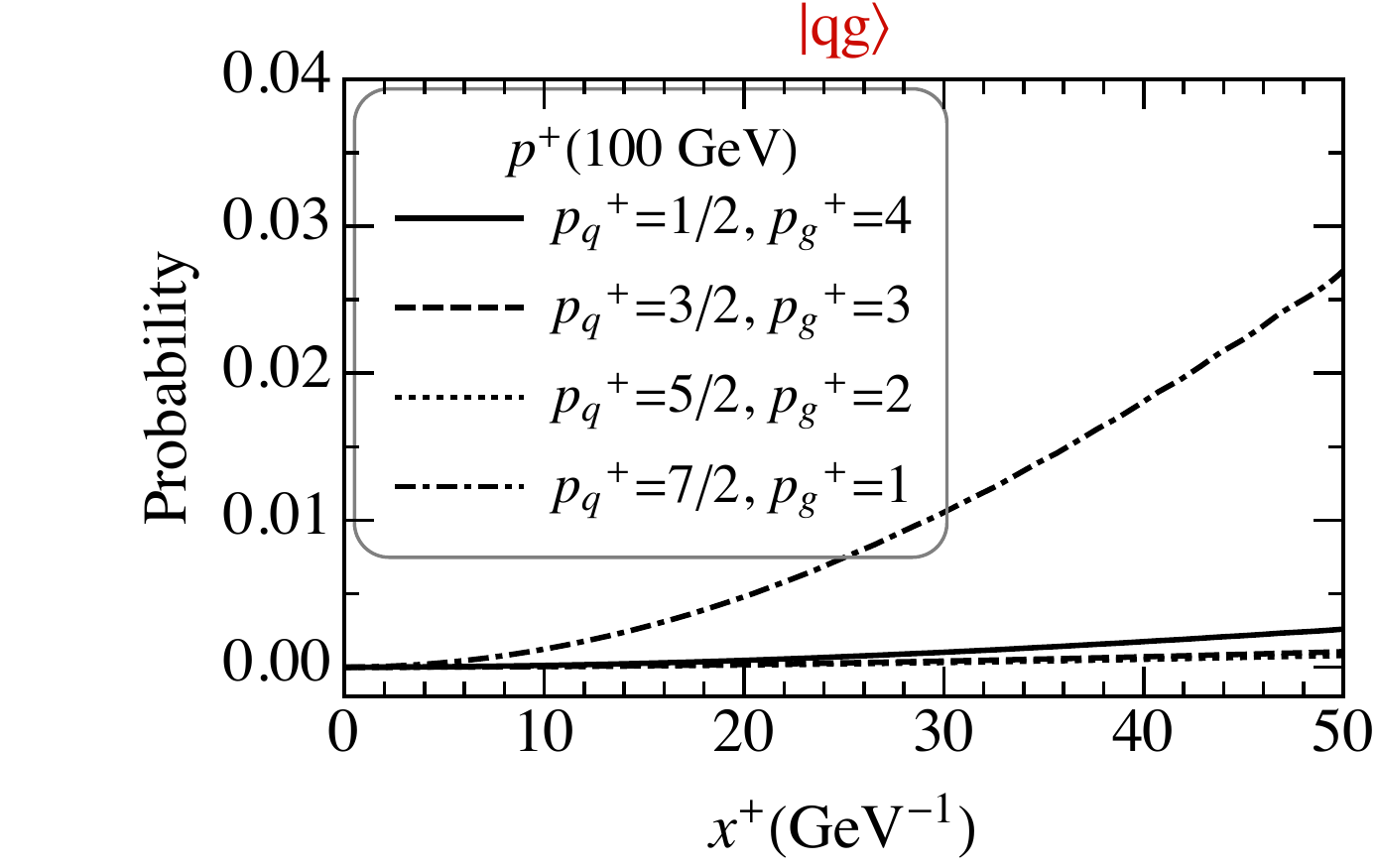}
  } 
  \caption{The evolution of the $p^+$ distribution in (a) the $\ket{q}$ sector and (b) the $\ket{qg}$ sector. Parameters in those panels: $L_\eta=50~\GeV^{-1}$, $N_\eta=8$, $L = 0.01 * 2\pi ~\GeV^{-1}$, $K_{\max}=4.5$, $m_g=0.1~\GeV$, $m_q=4.2~\GeV$, $N_\perp=4$, $L_\perp=50~\GeV^{-1}$, $g^2\tilde{\mu}=0.018~\GeV^{-3/2}$. 
  The initial state is a bare quark with $p_x=p_y=0$, $p^+= 4.5\times 100 ~\GeV$, spin up, and single color $c=1$.
  }
  \label{fig:ppl}
\end{figure}

In Fig.~\ref{fig:ppl}, we present the evolution of the $p^+$ distribution in the $\ket{q} + \ket{qg}$ Fock space. The decreasing probability of the single quark state and the emerging of the one-gluon-dressed quark indicate that the quark loses energy through the evolution, and that is via gluon emissions. 
\section{Conclusions}
We applied the tBLFQ formalism to the quark-nucleus scattering and study sub-eikonal effects from  non-perturbative aspects. 
We are able to access the wavefunction of the quark at intermediate time during the evolution. We foresee more applications to scattering processes in the near future.


\acknowledgments
This work was supported by the US Department of Energy (DOE) under Grant Nos. DE-FG02-87ER40371, DE-SC0018223 (SciDAC-4/NUCLEI), DE-SC0015376 (DOE Topical Collaboration in Nuclear Theory for Double-Beta Decay and Fundamental Symmetries). This research used resources of the National Energy Research Scientific Computing Center (NERSC), a U.S. Department of Energy Office of Science User Facility operated under Contract No. DE-AC02-05CH11231.
This work has been supported by the European Research Council (ERC) under the European Union’s Horizon 2020 research and innovation programme (grant agreement No ERC-2015-CoG-681707). The content of this article does not reflect the official opinion of the European Union and responsibility for the information and views expressed therein lies entirely with the authors.


\bibliographystyle{JHEP}
\bibliography{qA}

\end{document}